\documentclass[prb,aps,twocolumn,superscriptaddress,showpacs,preprintnumbers,amsmath,amssymb]{revtex4}
\usepackage{color}
\usepackage{graphicx}% Include figure files
\usepackage{ulem}
\usepackage{enumitem}
\def\textbf#1{\boldsymbol{#1}}

\begin{document}

\title{Anomalous dependence of the $c$-axis polarized Fe $B_{1g}$ phonon mode with Fe and Se concentrations in Fe$_{1+y}$Te$_{1-x}$Se$_{x}$}

\author{Y.J.~Um}
\affiliation{Max-Planck-Institut~f\"{u}r~Festk\"{o}rperforschung, Heisenbergstrasse~1, D-70569 Stuttgart, Deutschland}

\author{ A.~Subedi}
\affiliation{Max-Planck-Institut~f\"{u}r~Festk\"{o}rperforschung, Heisenbergstrasse~1, D-70569 Stuttgart, Deutschland}

\author{P.~Toulemonde}
\affiliation{Institut N\'{e}el, CNRS $\&$ UJF, 25 avenue des martyrs, F-38042 Grenoble cedex 09, France}

\author{A.Y.~Ganin}
\affiliation{Department of Chemistry, University of Liverpool, Liverpool, L69 7ZD, United Kingdom}

\author{L.~Boeri}
\affiliation{Max-Planck-Institut~f\"{u}r~Festk\"{o}rperforschung, Heisenbergstrasse~1, D-70569 Stuttgart, Deutschland}

\author{ M.~Rahlenbeck}
\affiliation{Max-Planck-Institut~f\"{u}r~Festk\"{o}rperforschung, Heisenbergstrasse~1, D-70569 Stuttgart, Deutschland}

\author{Y.~Liu}
\affiliation{Max-Planck-Institut~f\"{u}r~Festk\"{o}rperforschung, Heisenbergstrasse~1, D-70569 Stuttgart, Deutschland}

\author{ C.T.~Lin}
\affiliation{Max-Planck-Institut~f\"{u}r~Festk\"{o}rperforschung, Heisenbergstrasse~1, D-70569 Stuttgart, Deutschland}

\author{S.J.E.~Carlsson}
\affiliation{Institut N\'{e}el, CNRS $\&$ UJF, 25 avenue des martyrs, F-38042 Grenoble cedex 09, France}

\author{A. Sulpice}
\affiliation{Institut N\'{e}el, CNRS $\&$ UJF, 25 avenue des martyrs, F-38042 Grenoble cedex 09, France}
\affiliation{CRETA, CNRS $\&$ UJF, 25 avenue des martyrs, F-38042 Grenoble cedex 09, France}

\author{M.J.~Rosseinsky}
\affiliation{Department of Chemistry, University of Liverpool, Liverpool, L69 7ZD, United Kingdom}

\author{B.~Keimer}
\affiliation{Max-Planck-Institut~f\"{u}r~Festk\"{o}rperforschung, Heisenbergstrasse~1, D-70569 Stuttgart, Deutschland}

\author{M.~Le Tacon}
\affiliation{Max-Planck-Institut~f\"{u}r~Festk\"{o}rperforschung, Heisenbergstrasse~1, D-70569 Stuttgart, Deutschland}

\date{\today}

\begin{abstract}
We report on an investigation of the lattice dynamical properties in a range of Fe$_{1+y}$Te$_{1-x}$Se$_{x}$ compounds, with special emphasis on the $c$-axis polarized
vibration of Fe with $B_{1g}$ symmetry, a Raman active mode common to all families of Fe-based superconductors.
We have carried out a systematic study of the temperature dependence of this phonon mode as a function of Se $x$ and excess Fe $y$ concentrations.
In parent compound Fe$_{1+y}$Te, we observe an unconventional broadening of the phonon between room temperature and magnetic ordering temperature $T_N$.
The situation smoothly evolves toward a regular anharmonic behavior as Te is substituted for Se and long range magnetic order is replaced by
superconductivity.
Irrespective to Se contents, excess Fe is shown to provide an additional damping channel for the $B_{1g}$ phonon at low temperatures.
We performed Density Functional Theory \textit{ab initio} calculations within the local density approximation to calculate the phonon frequencies including magnetic polarization and Fe non-stoichiometry in the virtual crystal approximation. We obtained a good agreement with the measured phonon frequencies in the Fe-deficient samples, while the effects of Fe excess are poorly reproduced.
This may be due to excess Fe-induced local magnetism and low energy magnetic fluctuations that can not be treated accurately within these approaches.
As recently revealed by neutron scattering and muon spin rotation studies, these phenomena occur in the temperature range where
anomalous decay of the $B_{1g}$ phonon is observed and suggests a peculiar coupling of this mode with local moments and spin fluctuations in Fe$_{1+y}$Te$_{1-x}$Se$_{x}$.
\end{abstract}


%%%%%%%%%%%%%%%%%%%%%%%%%%%%INTRODUCTION%%%%%%%%%%%%%%%%%%%%%%%%%%%%%%%%%%%%%%%%%%%%%%%%%%%%%%%%
%%%%%%%%%%%%%%%%%%%%%%%%%%%%%%%%%%%%%%%%%%%%%%%%%%%%%%%%%%%%%%%%%%%%%%%%%%%%%%%%%%%%%%%%%%%%%%%%
\maketitle

\section{Introduction}
%introduction and motivation of FeTeSe
The recent discovery of superconductivity in F-doped LaFeAsO with $T_c$ of $\sim$ 26 K created a flurry of excitement in condensed matter research.~\cite{Kamihara08} Rapidly, numerous families of Fe-based superconductors such as $RE$FeAs(O$_{1-x}$F$_{x}$) (1111-family, $RE$ = rare earth), $M$Fe$_{2}$As$_{2}$ (122-family, $M$ = Ba, Ca, Sr, K, Cs ...), LiFeAs/NaFeAs (111-family) and Fe$_{1+y}$Te$_{1-x}$Se$_{x}$ (11-family)~\cite{Takahashi08,Ren08,Rotter08,Tapp08,Hsu08, Yeh_EPL2008,Fang08} have been found and investigated.
Although the Fe$_{1+y}$Te$_{1-x}$Se$_{x}$ chalcogenides share with their pnictogen-based cousins similar structure, based on planar layers of edge sharing [Fe(Se,Te)]$_4$ tetrahedra, some significant differences were rapidly revealed.
Among them are the large magnetic moment ($\sim$ 2.0-2.5 $\mu_B$) and the double stripe magnetic ordering of the parent Fe$_{1+y}$Te compound.~\cite{Fruchart_1975,Li09} Magnetic stripes are found in other Fe-based parent compounds such as BaFe$_2$As$_2$ or LaFeAsO stripes in pnictides, but they are rotated by 45$^{\circ}$ around the $c$-axis with respect to Fe$_{1+y}$Te,~\cite{Bao09,Li09, Ma_PRL2009, Han_PRL2009} with significantly lower magnetic moments on Fe in these compound (typically 0.4-1.0 $\mu_B$/iron~\cite{delaCruz_Nature2008, Singh08}).

The magnetic ordering is accompanied by a structural transition ($T_N \sim 67 K$),~\cite{Fruchart_1975} which can be progressively suppressed on substitution of isovalent Se at the Te 2b Wyckoff position. This also results in emergence of superconductivity with a $T_c^{max} \sim$ 14 K at ambient pressure at the optimum doping.~\cite{Hsu08, Yeh_EPL2008}
However, numerous reports have recently demonstrated that the excess of interstitial Fe between the chalcogenide layers, even after doping with Se, could affect both superconducting and magnetic properties,  \textit{e.g.} suppression of transition temperature $T_c$ or lowering shielding fraction, as well as leading to the appearance of weakly localized magnetic states.~\cite{Martinelli_PRB2010,Khasanov09,Zhang09,Wen09,Liu09,Bendele10,Xu_PRB2011}

%A proposed explanation for this rotation is based on changes in the Fermi surface nesting properties induced by small excess iron in Fe$_{1+y}$Te ($0.02 \lesssim y \lesssim 0.15$)~\cite{Han_PRL2009} in interstitial 2c Wyckoff position (\textit{i.e.} between the Fe-Te planes).
%This emphasize the importance of this Fe excess, recently shown to affect both superconducting and magnetic properties, leading for instance to decrease of superconducting transition temperature $T_c$ and to the apparition of weakly localized magnetic states~\cite{Martinelli_PRB2010,Khasanov09,Zhang09,Wen09,Liu09,Bendele10,Xu_PRB2011}.

%motivation of Raman scattering
Raman spectroscopy is an ultimate noninvasive tool and can allow systematic studies of the temperature dependence of the phonon spectrum as function of Se and excess of interstitial Fe. For example, Raman spectroscopy has proved useful in the understanding of the structural, magnetic and electronic properties of superconducting pnictides. ~\cite{Hanh_PRB2009,Litvinchuk08,Reznik09,Rahlenbeck09,Chauviere09, Fukuda_JPCS2008, LeTacon_PRB2008, Le Tacon09,Choi10, Gallais08, Choi_PRB2008, Fukuda_PRB2011, Um_PRB2012,Litvinchuk_PRB2011}
Available Raman data on single crystals of Fe$_{1+y}$Te$_{1-x}$Se$_{x}$ is to date limited.
%Little work has been, however, performed on well-characterized single crystals of Fe$_{1+y}$Te$_{1-x}$Se$_{x}$ ~\cite{Xia09,Okazaki11, Gnezdilov}.
In refs.~\onlinecite{Xia09} and~\onlinecite{Okazaki11}, comparison between Se-substituted and parent Fe$_{1+y}$Te samples is made, but neither temperature dependence nor influence of the excess iron concentration on the lattice dynamics are discussed.
In ref.~\onlinecite{Gnezdilov}, the authors study the evolution with temperature of the Raman spectra of Fe$_{1.05}$Te, with a particular emphasis on the anomalously large lineshape of the Te $A_{1g}$ phonon. It is argued to originate from a peculiar spin-orbital frustration effect, that leaves unaffected the $B_{1g}$ phonon after symmetry considerations.

In this paper, we focus mainly on this Raman active $c$-axis polarized optical phonon in Fe$_{1+y}$Te$_{1-x}$Se$_{x}$, which is common to all the iron-based superconductors families, and discuss its evolution with temperature for various Fe, $y$, and Se content, $x$. We can demonstrate that contrarily to the $A_{1g}$ mode, the observed temperature dependence is strongly affected by composition. To some extent, the phonon behavior through the various phases transitions (depending of the Se content) is consistent with those of the $c$-axis polarized Fe modes reported in the other families of Fe-based superconductors (\textit{e.g.} 122 and 111). The narrowing of the phonon lineshape through the magnetic transition of the parent compound or the absence of renormalization through the superconducting one are for instance reported.
On the other hand, we also show that in some specific conditions, the behavior of the $B_{1g}$ mode seriously deviates from the aforementioned one. The phonon linewidth shows an anomalous broadening in the paramagnetic state of Fe$_{1+y}$Te parent compounds and an unusually strong dependence with the Se concentration in the doped compounds. Further anomalies, indicative of additional decay channels, are found when increasing the concentration of excess Fe $y$.

To try to get some insights about the influence of excess iron on the Raman phonon, we have carried out a Density Functional Theory (DFT) \textit{ab initio} calculation within the local density approximation (LDA) of the phonon frequencies, including the effects of magnetism and Fe nonstoichiometry in the virtual crystal approximation (VCA). The measured frequencies are in good agreement with the predicted ones, including a softening with increasing Fe content in Fe deficient samples, but the effects of Fe nonstoichiometry are poorly reproduced.

Recent studies of magnetic properties have revealed that excess Fe induces local magnetism and low energy magnetic fluctuations,~\cite{Xu_PRB2011, Stock} but these can not be treated within our DFT approaches.
Generally speaking, whenever low-energy magnetic fluctuations are at play, the $B_{1g}$ phonon behavior deviates from the conventional harmonic picture and cannot be reproduced within our theoretical framework. This suggests that the reported anomalies originate from coupling between the $B_{1g}$ mode and these excitations.

%%%%%%%%%%%%%%%%%%%%%%%%%%%%%%%%EXPERIMENT%%%%%%%%%%%%%%%%%%%%%%%%%%%%%%%%%%%%%%%%%%%%%%%%%%%%%%%
%%%%%%%%%%%%%%%%%%%%%%%%%%%%%%%%%%%%%%%%%%%%%%%%%%%%%%%%%%%%%%%%%%%%%%%%%%%%%%%%%%%%%%%%%%%%%%%%%
\section{Experimental Details}

\begin{table}
\caption{\label{tab:TableSample}A summary of the chemical compositions obtained by EDX and characteristic transitions temperatures of the various Fe$_{1+y}$Te$_{1-x}$Se$_{x}$ samples used in this study.}
\begin{ruledtabular}
\begin{tabular}{ccc}
Sample Composition & T$_{N}$ &T$_{c}$\\
\hline
Fe$_{1.02}$Te  &   67 K &~~~---~~\\
Fe$_{1.09}$Te  &   65 K &~~~---~~ \\
\hline
%Fe$_{1+y}$Te$_{1-x}$Se$_{x}$       & T$_{c}$  \\
Fe$_{1.00}$Te$_{0.78}$Se$_{0.22}$    &  ---~~ & 11.5 K  \\
Fe$_{0.99}$Te$_{0.69}$Se$_{0.31}$   &   ---~~ & 11 K    \\
Fe$_{0.98}$Te$_{0.66}$Se$_{0.34}$   &   ---~~ & 10.5 K  \\
Fe$_{0.95}$Te$_{0.56}$Se$_{0.44}$  &   ---~~ & 14 K    \\

\hline
Fe$_{1.05}$Te$_{0.58}$Se$_{0.42}$ & ---  & 11.5 K \\
Fe$_{1.08}$Te$_{0.73}$Se$_{0.27}$ & ---  & 9 K \\
\end{tabular}
\end{ruledtabular}
\end{table}

\begin{figure}
\includegraphics[width=\linewidth]{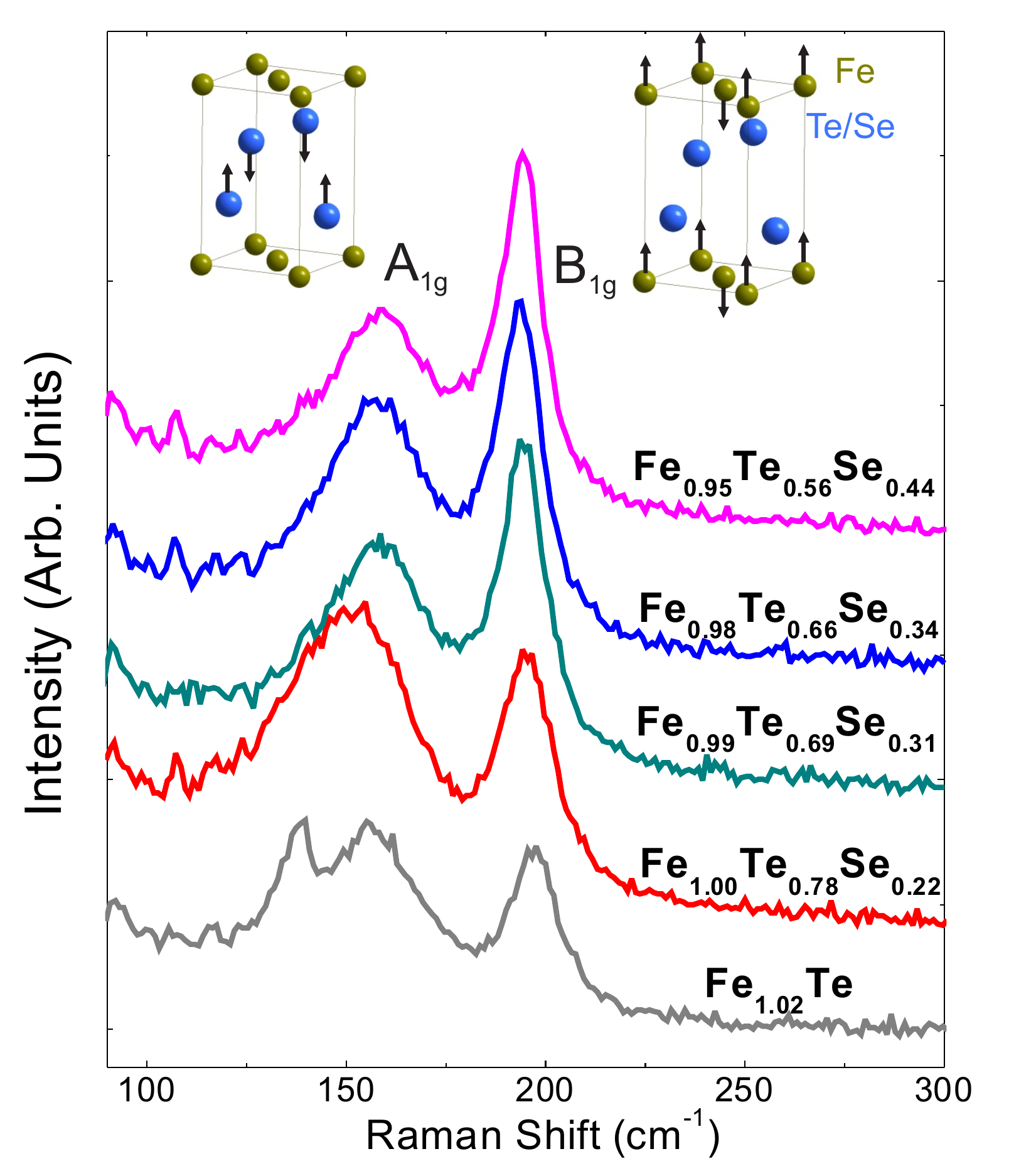}
\caption{(Color online) Room temperature Raman spectra of the Fe$_{1.02}$Te, Fe$_{1.00}$Te$_{0.78}$Se$_{0.22}$, Fe$_{0.99}$Te$_{0.69}$Se$_{0.31}$, Fe$_{0.98}$Te$_{0.66}$Se$_{0.34}$, and Fe$_{0.95}$Te$_{0.56}$Se$_{0.44}$ samples (see Table~\ref{tab:TableSample}). Spectra have been shifted vertically for clarity. Eigendisplacements corresponding to the $A_{1g}$(Te/Se) and $B_{1g}$(Fe) are represented schematically.}
\label{FigEigen}
\end{figure}

%Sample characterization
In order to discriminate the effects of Fe excess and of Se substitution, we have studied different groups of crystals, listed in Table~\ref{tab:TableSample}. In the first group, no Se was present and only the Fe concentration was changed Fe$_{1+y}$Te (Fe$_{1.02}$Te, Fe$_{1.09}$Te). In the second group we changed the Se concentration while keeping the Fe concentration as close from 1 as possible (Fe$_{1.00}$Te$_{0.78}$Se$_{0.22}$, Fe$_{0.99}$Te$_{0.69}$Se$_{0.31}$, Fe$_{0.98}$Te$_{0.66}$Se$_{0.34}$  and Fe$_{0.95}$Te$_{0.56}$Se$_{0.44}$). Finally, we also studied Se-substituted samples containing sizable excess iron (Fe$_{1.05}$Te$_{0.58}$Se$_{0.42}$, Fe$_{1.08}$Te$_{0.73}$Se$_{0.27}$).
Fe$_{1+y}$Te$_{1-x}$Se$_{x}$ single crystals were grown in sealed quartz tube as described elsewhere~\cite{Liu10, Klein_PRB2010, Tamai_PRL2010, Gresty_JACS2009}.
The chemical compositions listed in Table~\ref{tab:TableSample} were determined using energy dispersive x-ray spectroscopy (EDX), and the antiferromagnetic (AF) and superconducting (SC) transition temperatures $T_N$ and $T_c$ were measured by use of a superconducting quantum interference device magnetometer.

\begin{figure*}
\includegraphics[width=0.9\linewidth]{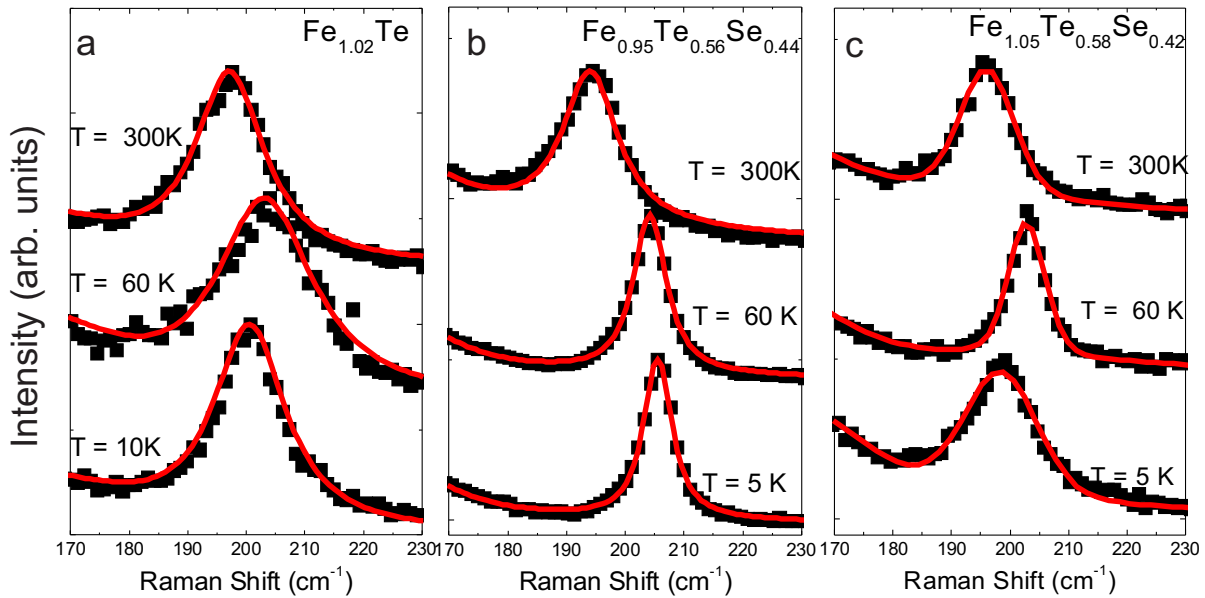}
\caption{(Color online) a) $B_{1g}$ phonon of the Fe$_{1.02}$Te sample for selected temperatures (room temperature, $T \sim T_N$ and base temperature). Black squares are the data, and the red line is the fit following the procedure described in the text. Phonon intensity have been normalized and the spectra have been shifted vertically for clarity. b) Same plot for the Fe$_{0.95}$Te$_{0.56}$Se$_{0.44}$ sample. c) same plot for the Fe$_{1.05}$Te$_{0.58}$Se$_{0.42}$ sample.}
\label{Fig1}
\end{figure*}

%Raman scattering
All Raman light scattering experiments were performed on freshly cleaved surface of Fe$_{1+y}$Te$_{1-x}$Se$_{x}$ single crystals. All the samples were mounted in a helium-flow cryostat allowing measurements between 5 K and room temperature. Spectra were taken in backscattering geometry through a JobinYvon LabRam 1800 single grating spectrometer equipped with a razor-edge filter and a Peltier-cooled charge-coupled-device camera.
We used a linearly polarized He$^{+}$/Ne$^{+}$ mixed gas laser with $\lambda$ = 632.817 nm for excitation.
The laser beam was focused through a 50$\times$ microscope objective to $\sim$5 $\mu$m diameter spot on the sample surface. The power of the incident laser was kept less than 1 mW to avoid laser-induced heating. In order to determine the precise frequency of phonons for each temperature, Neon emission lines were recorded between each measurements. For data analysis, all phonon peaks were fitted by Lorentzian profiles, convoluted with the spectrometer resolution function (a Gaussian line of 2 cm$^{-1}$ full width at half maximum (FWHM)).

%%%%%%%%%%%%%%%%%%%%%%%%%%%%%RESULTS%%%%%%%%%%%%%%%%%%%%%%%%%%%%%%%%
%%%%%%%%%%%%%%%%%%%%%%%%%%%%%%%%%%%%%%%%%%%%%%%%%%%%%%%%%%%%%%%%%%%%

\section{Experimental Results}
\label{Results}
\subsection{Influence of Se doping}
\label{ResultsSe}

\begin{figure*}
\includegraphics[width=1\linewidth]{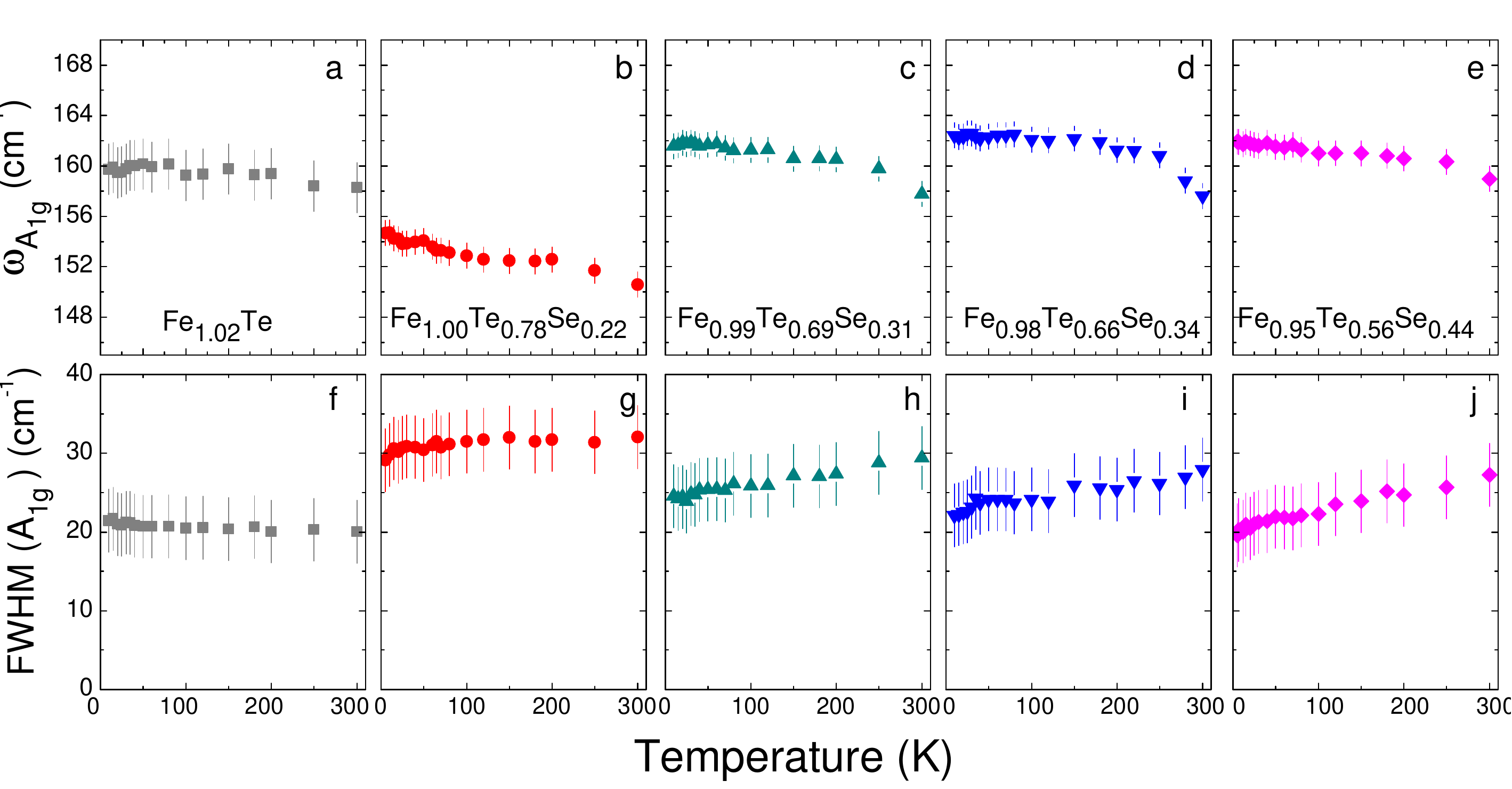}
\caption{(Color online) Upper panel: Temperature dependence of the $A_{1g}$(Te/Se) mode frequency of the a) Fe$_{1.02}$Te, b) Fe$_{1.00}$Te$_{0.78}$Se$_{0.22}$, c) Fe$_{0.99}$Te$_{0.69}$Se$_{0.31}$, d) Fe$_{0.98}$Te$_{0.66}$Se$_{0.34}$, and e) Fe$_{0.95}$Te$_{0.56}$Se$_{0.44}$ samples. Lower panel: Temperature dependence of the $A_{1g}$(Te/Se) mode FWHM of the f) Fe$_{1.02}$Te, g) Fe$_{1.00}$Te$_{0.78}$Se$_{0.22}$, h) Fe$_{0.99}$Te$_{0.69}$Se$_{0.31}$, i) Fe$_{0.98}$Te$_{0.66}$Se$_{0.34}$, and j) Fe$_{0.95}$Te$_{0.56}$Se$_{0.44}$ samples.}
\label{Fig_A1g}
\end{figure*}

\begin{figure*}
\includegraphics[width=1\linewidth]{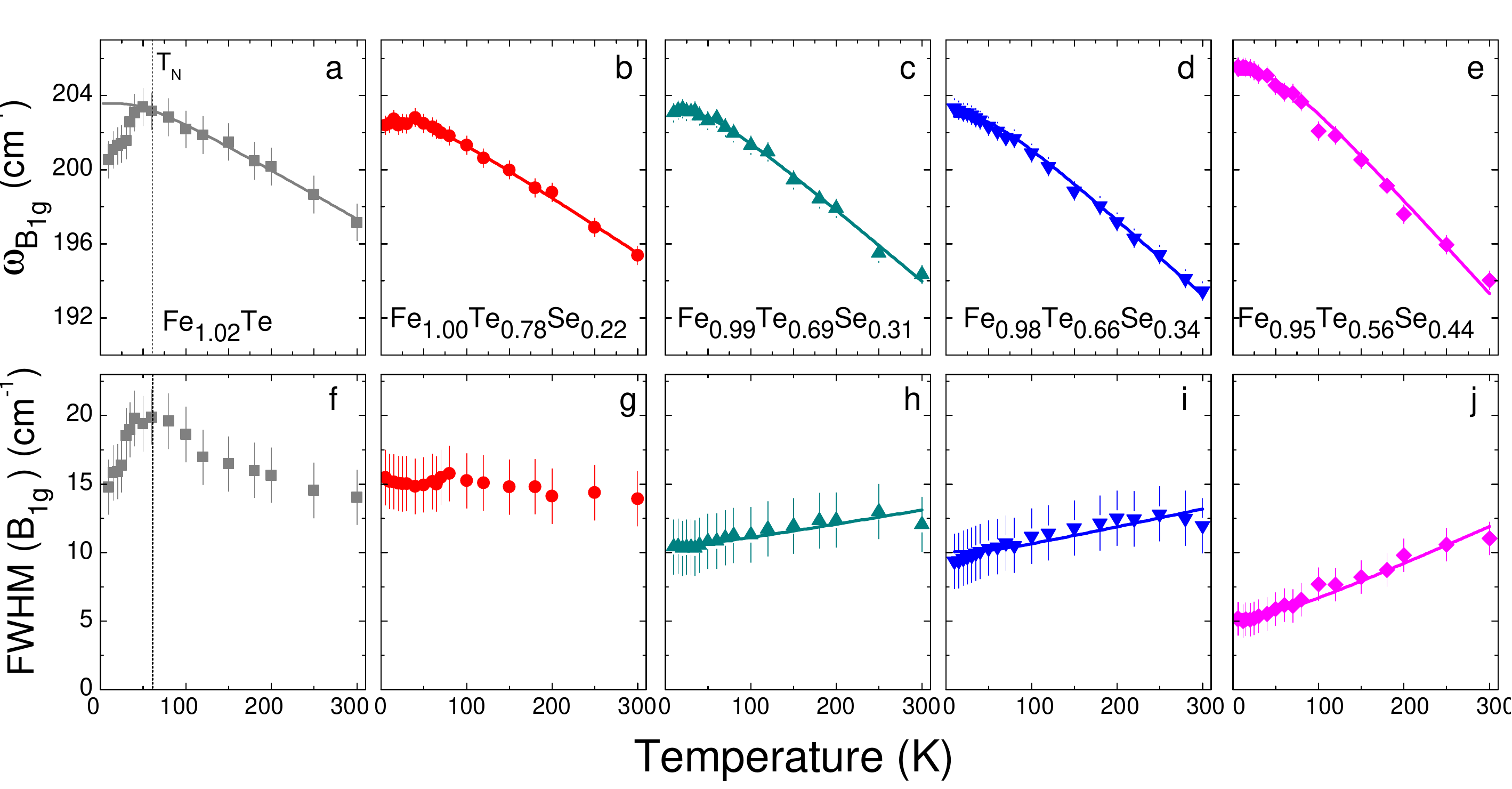}
\caption{(Color online) Upper panel: Temperature dependence of the $B_{1g}$(Fe) mode frequency of the a) Fe$_{1.02}$Te, b) Fe$_{1.00}$Te$_{0.78}$Se$_{0.22}$, c) Fe$_{0.99}$Te$_{0.69}$Se$_{0.31}$, d) Fe$_{0.98}$Te$_{0.66}$Se$_{0.34}$, and e) Fe$_{0.95}$Te$_{0.56}$Se$_{0.44}$ samples. Lower panel: Temperature dependence of the $B_{1g}$(Fe) mode full-width-half-maximum of the f) Fe$_{1.02}$Te, g) Fe$_{1.00}$Te$_{0.78}$Se$_{0.22}$, h) Fe$_{0.99}$Te$_{0.69}$Se$_{0.31}$, i) Fe$_{0.98}$Te$_{0.66}$Se$_{0.34}$, and j) Fe$_{0.95}$Te$_{0.56}$Se$_{0.44}$ samples. Solid lines are fits of the temperature dependence of the $B_{1g}$ phonon frequency and linewidth in the various samples using the anharmonic model described in the text.}
\label{Fig_B1g}
\end{figure*}

In Fig.~\ref{FigEigen} we show the Raman spectrum measured at room temperature on the Se-free Fe$_{1.02}$Te sample, together with the Fe$_{1.00}$Te$_{0.78}$Se$_{0.22}$, Fe$_{0.99}$Te$_{0.69}$Se$_{0.31}$, Fe$_{0.98}$Te$_{0.66}$Se$_{0.34}$  and Fe$_{0.95}$Te$_{0.56}$Se$_{0.44}$ superconducting samples that all have a Fe stoichiometry close to 1. From symmetry consideration, one expects four Raman active modes:  $A_{1g}$(Te/Se), $B_{1g}$(Fe), $E_{g}$(Te), $E_{g}$(Fe).
Raman measurement have been performed in backscattering geometry, with the incident light polarization along the $a$-axis of the single crystal.
An analyzer has been used to check the phonon selection rules, but most of the measurements presented here have been performed without in order to maximize the phonon peak intensities.

\begin{figure}
\includegraphics[width=0.7\linewidth]{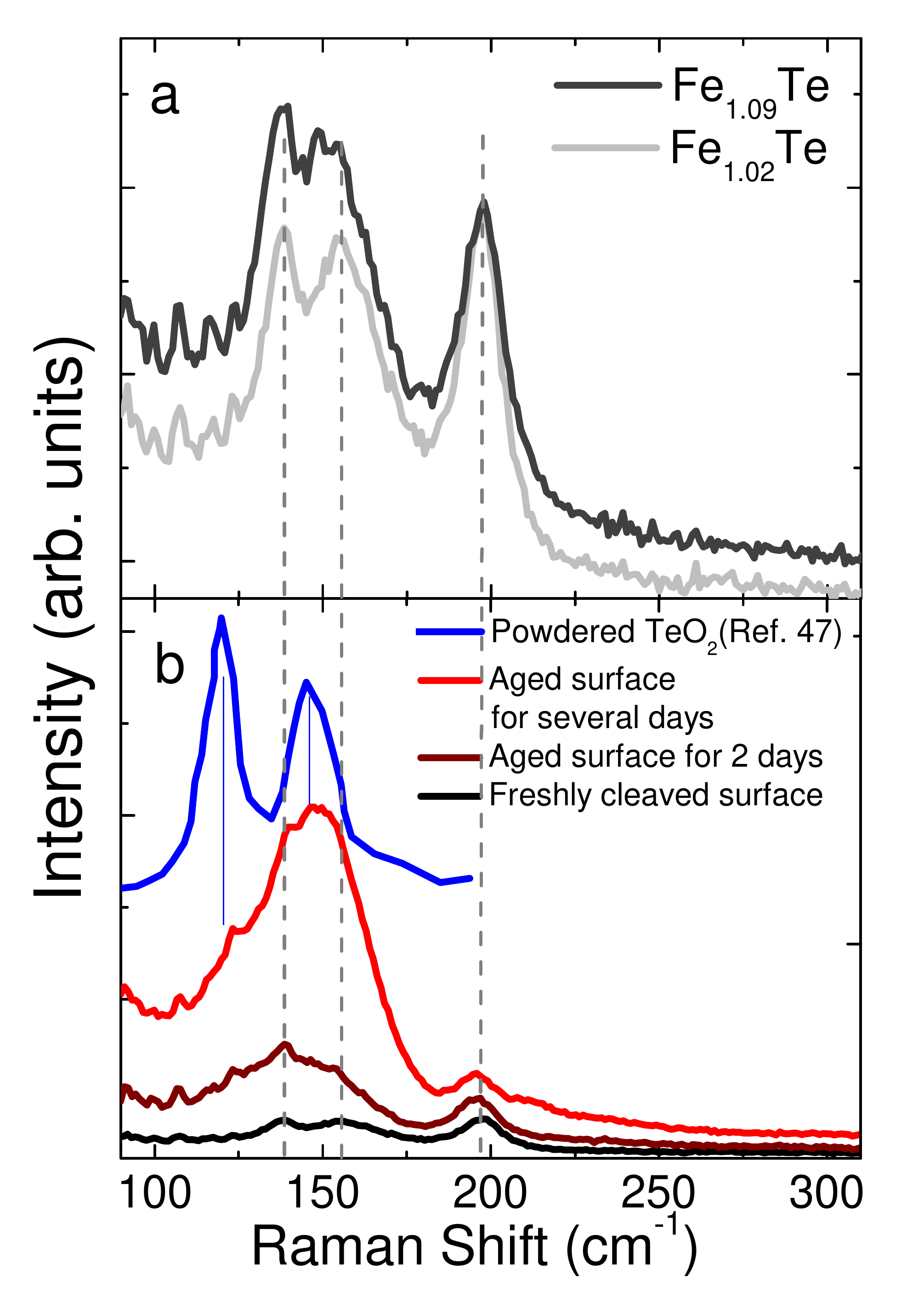}
\caption{(Color online) a) Room temperature Raman spectra of the parent Fe$_{1.02}$Te and Fe$_{1.09}$Te samples (vertically shifted for clarity). b) Example of the aging effect on the parent single crystals. The spectrum of powdered TeO$_2$ from ref.~\cite{Pine71} has been added for comparison.}
\label{Fig3a}
\end{figure}

In this scattering geometry, only the $A_{1g}$(Te/Se) and $B_{1g}$(Fe) modes, sketched in Fig.~\ref{FigEigen}, are present. As seen in Fig.~\ref{FigEigen}, these modes are found at $\sim$ 155 cm$^{-1}$ and $\sim$ 197 cm$^{-1}$ at room temperature, respectively, in agreement with previous reports~\cite{Xia09,Okazaki11,Gnezdilov}. These two peaks are much broader than in any of the other iron pnictides~\cite{Rahlenbeck09, Chauviere09, Gallais08}. The large $A_{1g}$ mode linewidth ($\sim$ 20 cm$^{-1}$ at room temperature in Fe$_{1.02}$Te, almost 3 times larger than the $A_{1g}$ As mode in BaFe$_2$As$_2$~\cite{Rahlenbeck09}) has  been attributed to spin-orbital frustration effects.~\cite{Gnezdilov} In the parent Fe$_{1.02}$Te single crystal, an additional peak was observed around 136 cm$^{-1}$. The origin of this mode remains unclear. It is temperature independent (and cannot, therefore, be attributed to the lowering of the crystal symmetry induced by the structural transition), and has been observed irrespective of the Fe excess concentration (see Fig.~\ref{Fig3a}-a). It is not observed in the Se-rich compounds (see Fig.~\ref{FigEigen}).
In contrast with the results of Xia \textit{et al.} in ref.~\onlinecite{Xia09}, claiming the disappearance of the $A_{1g}$ mode in the Raman spectra of FeTe$_{0.92}$ and Fe$_{1.03}$Te$_{0.7}$Se$_{0.3}$ with increasing Se concentration, the mode is clearly visible in all the investigated compounds.

The frequency of the two phonons is weakly dependent on Se contents: at the lowest recorded temperatures (5 K) it remains essentially constant for Se contents between 22$\%$ and 34$\%$ (Fe$_{1.00}$Te$_{0.78}$Se$_{0.22}$ and Fe$_{0.98}$Te$_{0.66}$Se$_{0.34}$), while a small hardening ($\sim$ 2 cm$^{-1}$) is observed for the sample with 44$\%$ of Se (Fe$_{0.95}$Te$_{0.56}$Se$_{0.44}$).
This latter effect may be caused by the significant Fe deficiency in the Fe$_{0.95}$Te$_{0.56}$Se$_{0.44}$ sample (see Secs.~\ref{DFTcalc} and ~\ref{disc}). Within our experimental errors, this seems to be also the case for the $A_{1g}$ phonon, as trivially expected from the substitution of Te with lighter Se (the mode frequency going, in first approximation, as $M^{-1/2}$, with $M$ denoting the reduced mass of the considered oscillator). The only noticeable exception, is the Fe$_{1.00}$Te$_{0.78}$Se$_{0.22}$ sample, where the $A_{1g}$ mode is broader and softer than in any other compounds. This may originate from an overlapping of the $A_{1g}$ mode with the 136 cm$^{-1}$ peak observed in the parent compounds. It seems that for this particular doping level the modes energies are still separated enough to cause an apparent broadening and shift to lower frequency of the total envelope, but not enough to allow to resolve them individually.

In Figs.~\ref{Fig1}-a and -b, we show details of the fitting for the Fe$_{1.02}$Te and Fe$_{0.95}$Te$_{0.56}$Se$_{0.44}$ samples, that illustrates one of the main finding of our study, \textit{i.e.} the strong changes in the temperature dependence of the $B_{1g}$ phonon on doping. In the undoped Fe$_{1.02}$Te sample, the $B_{1g}$ mode hardens and broaden with decreasing temperature down to $T_N$, and then softens and narrows down to base temperature. For the Fe$_{0.95}$Te$_{0.56}$Se$_{0.44}$ sample with the highest Se concentration, we observed a continuous hardening and narrowing of the mode the whole way down to 5 K.
In Figs.~\ref{Fig_A1g} and~\ref{Fig_B1g}, respectively, we report the full temperature dependence of the frequency of the two $c$-axis polarized modes for the five samples with Fe concentration close to 1. The temperature dependencies of the linewidths for the same samples are given in the lower panels of these figures.

\begin{figure}
\includegraphics[width=\linewidth]{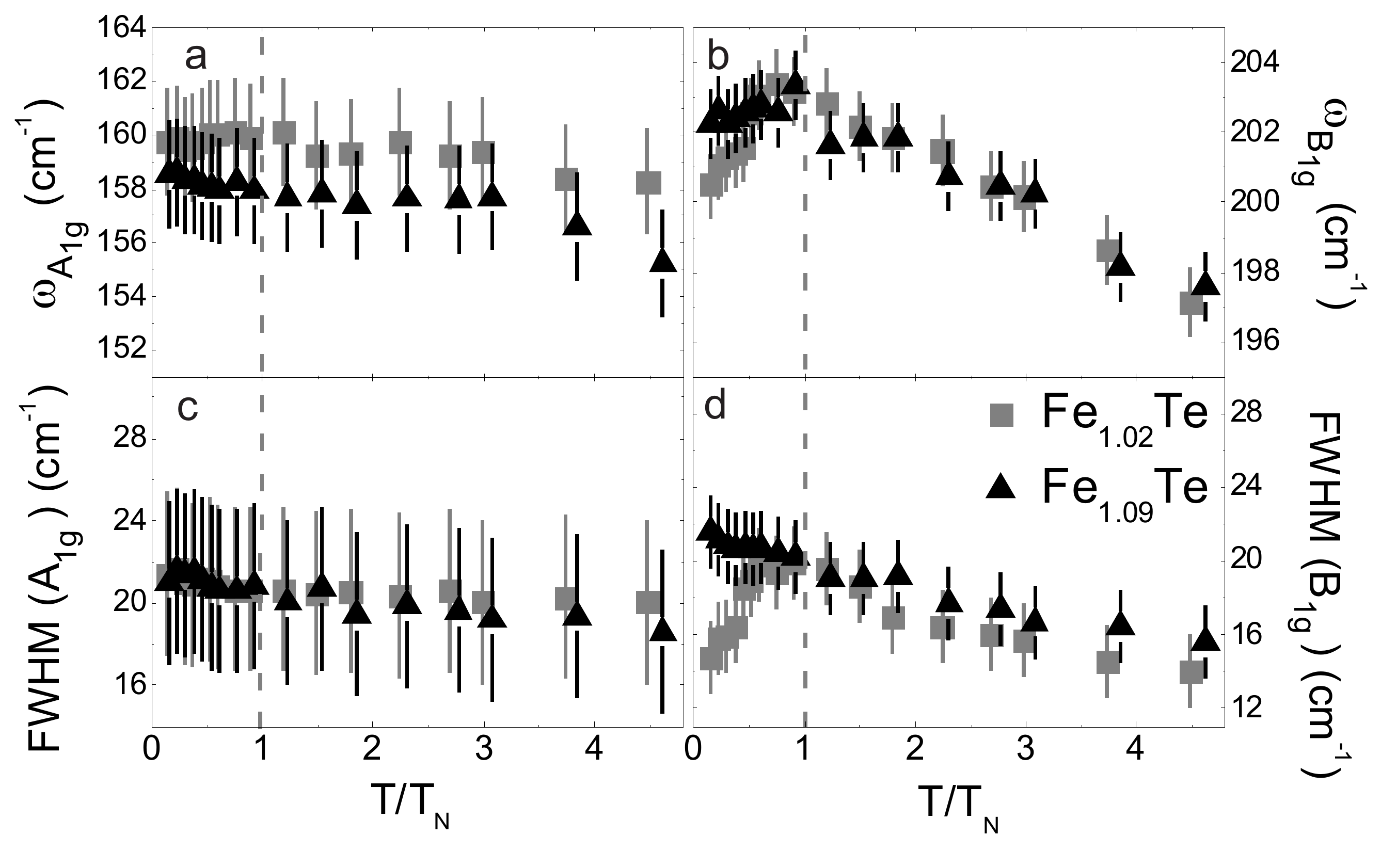}
\caption{(Color online) a) and b) Temperature dependence of the frequency the $A_{1g}$ and $B_{1g}$ modes in these samples. c) and d) Temperature dependence of the linewidth the $A_{1g}$ and $B_{1g}$ modes in these samples.}
\label{Fig3b}
\end{figure}

While the temperature is decreased, a hardening of both the $A_{1g}$ and $B_{1g}$ modes in all the systems is observed, as expected from the lattice contraction. No noticeable differences between the samples are seen.
As we go through the magnetic transition in the Fe$_{1.02}$Te parent compound, a clear softening of the $B_{1g}$ mode is observed (see also Fig.~\ref{Fig3b}), while no changes across $T_c$ occur in the samples containing Se. Within our error bars, the $A_{1g}$ mode frequency remains essentially unaffected by these transitions.
As shown in Fig.~\ref{Fig_A1g} weak narrowing of the $A_{1g}$ line with decreasing temperature is observed for Fe$_{0.99}$Te$_{0.69}$Se$_{0.31}$, Fe$_{0.98}$Te$_{0.66}$Se$_{0.34}$  and Fe$_{0.95}$Te$_{0.56}$Se$_{0.44}$ samples, while its broad linewidth remains essentially temperature independent in the Fe$_{1.02}$Te and Fe$_{1.00}$Te$_{0.78}$Se$_{0.22}$ sample. In parallel to this, an unusual evolution on Se doping of the temperature dependence of the $B_{1g}$ mode linewidth can be see in the lower panels of Fig.~\ref{Fig_B1g}.
Starting from the almost half-doped Fe$_{0.95}$Te$_{0.56}$Se$_{0.44}$ compound, a conventional behavior is observed: as in most of the materials, this mode narrows with decreasing temperature (the phonon width is inversely proportional to its lifetime, which is expected to increase as the phonon-phonon interaction is reduced when decreasing the temperature~\cite{Klemens66,Menendez84}). Decreasing Se concentration toward the parent Fe$_{1.02}$Te sample, a smooth evolution from this regular behavior is observed: in Fe$_{0.99}$Te$_{0.69}$Se$_{0.31}$ and Fe$_{0.98}$Te$_{0.66}$Se$_{0.34}$  samples only a weak narrowing of the phonon is observed between room and base temperature, while in Fe$_{1.00}$Te$_{0.78}$Se$_{0.22}$ and Fe$_{1.02}$Te the phonon essentially broadens as temperature is decreased. Below the magnetic ordering transition in the Fe$_{1.02}$Te sample, in addition to the softening mentioned above, a narrowing of the $B_{1g}$ phonon is observed, in agreement with recent report on Fe$_{1.05}$Te.~\cite{Gnezdilov}

The temperature dependence of both frequency and FWHM of the $B_{1g}$ phonon of the Fe$_{0.99}$Te$_{0.69}$Se$_{0.31}$, Fe$_{0.98}$Te$_{0.66}$Se$_{0.34}$  and Fe$_{0.95}$Te$_{0.56}$Se$_{0.44}$ samples can been well fitted assuming a symmetric anharmonic decay of this optical phonon, \textit{i.e.} decay into two acoustic modes with identical frequencies and opposite momenta,~\cite{Klemens66, Menendez84}
\begin{equation}\label{e1}
\omega_{ph}(\textit{T}) = \omega_{0} - C\bigg[1+\frac{2}{e^{\frac{\hbar\omega_{0}}{2k_{B}T}}-1}\bigg]
\end{equation}
\begin{equation}\label{e2}
\Gamma_{ph}(\textit{T}) = \Gamma_{0} + \Gamma\bigg[1+\frac{2}{e^{\frac{\hbar\omega_{0}}{2k_{B}T}}-1}\bigg]
\end{equation}
where  $C$ and $\Gamma$ are positive constants, $\omega_{0}$ is the bare phonon frequency, and $\Gamma_{0}$ a residual (temperature-independent) linewidth originating from sample imperfections or electron-phonon interactions.
The fitting parameters for these three samples are summarized in Table~\ref{tab:TableFitB1g}.
For the Fe$_{1.02}$Te and Fe$_{1.00}$Te$_{0.78}$Se$_{0.22}$ samples, as the FWHM increases with decreasing temperature, we can simply not use the latter expression to fit the experimental data.

\begin{table}
\caption{\label{tab:TableFitB1g}Fitting parameters for the temperature dependence of the $B_{1g}$ phonon linewidth in Fe$_{0.99}$Te$_{0.69}$Se$_{0.31}$, Fe$_{0.98}$Te$_{0.66}$Se$_{0.34}$  and Fe$_{0.95}$Te$_{0.56}$Se$_{0.44}$ samples.}
\begin{ruledtabular}
\begin{tabular}{cccc}
Sample               &        $\omega_0$ (cm$^{-1}$)      & $\Gamma_0$ (cm$^{-1}$) & $\Gamma$ (cm$^{-1}$) \\
\hline
Fe$_{0.99}$Te$_{0.69}$Se$_{0.31}$               &    203.9         &  9.9 & 0.78 \\
Fe$_{0.98}$Te$_{0.66}$Se$_{0.34}$                &     203.8        &  9.8 & 0.78 \\
Fe$_{0.95}$Te$_{0.56}$Se$_{0.44}$               &       205.6      &  3.44 & 2.1 \\
\end{tabular}
\end{ruledtabular}
\end{table}

\subsection{Effect of the iron excess}
\subsubsection{Undoped compounds}
\label{Exp_Fe_undoped}
Before discussing the possible origin of this unusual evolution of the temperature dependence of the $B_{1g}$ mode FWHM, it is interesting to discuss its dependence with the concentration of Fe excess.
We show in Fig.~\ref{Fig3a}-a the raw Raman data obtained at room temperature on the two Se-free samples Fe$_{1.02}$Te and Fe$_{1.09}$Te where the excess iron concentration has been measured to be 2\%, and 9\%. No strong differences can be found at first glance.
The strong aging effect reported in ref.~\cite{Xia09} has also been observed in our Fe$_{1+y}$Te samples, as seen in Fig.~\ref{Fig3a}-b, as well as on the Se-substituted samples (not shown here).
The origin of this strong Raman signal is not clear. In ref.~\cite{Xia09}, the authors attribute it to the formation of amorphous Te as a decomposition product of Fe$_{1+y}$Te$_{1-x}$Se$_{x}$ on the basis of earlier reports~\cite{Pine71,Pine72}. In particular, in Fig. 2-d from ref.~\cite{Pine71}, the authors say that the room temperature spectra of amorphous Te is plotted, but in a note added in proof, they recognize that it rather originates from tetragonal TeO$_2$ (the Raman spectra of single crystalline paratellurite TeO$_2$ reported in ref.~\cite{Pine72} shows a much complex structure).
We have added these data to our Fig.~\ref{Fig3a}-a and found them not inconsistent with the measurements from our aged surface. The main differences are the relative intensities and widths of the two features of TeO$_2$ at 120 and 145 cm$^{-1}$ that may originate either from different texturing and strain of TeO$_2$ and/or from the presence of Fe in the decomposition product.

The excess iron induces a small softening of the $A_{1g}$ mode (Fig.~\ref{Fig3b}-a) but does not affect its already broad linewidth (Fig.~\ref{Fig3b}-c). No changes in the temperature dependence are seen.
Similarly, the case of the $B_{1g}$ mode appears to be more interesting. In the excess-Fe rich sample Fe$_{1.09}$Te, a small softening through the magnetic transition can still be observed (Fig.~\ref{Fig3b}-b), but the signature of this transition in the FWHM is clearly suppressed (Fig.~\ref{Fig3b}-d).
We note finally that the unusual broadening of the mode with decreasing temperature in the parent compounds is observed irrespective to the measured excess iron concentration (fig.~\ref{Fig3b}-d), ruling out a disorder origin for this phenomena.

\subsubsection{Se-substituted compounds}
\begin{figure}
\includegraphics[width=\linewidth]{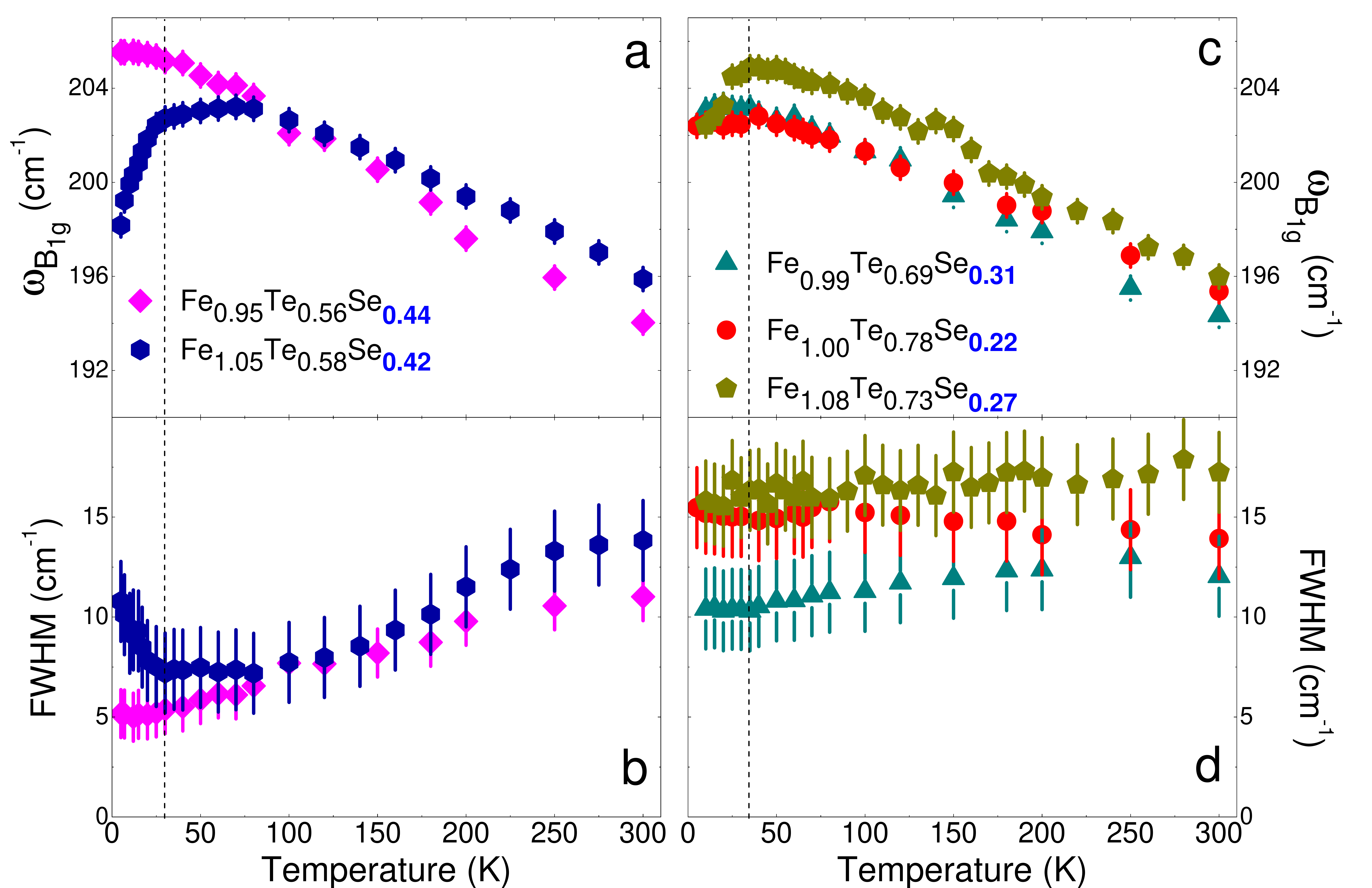}
\caption{(Color online) a) and c) Effect of the Fe excess on the temperature dependence of the $B_{1g}$ mode frequency in Se-substituted samples. b) and d)  Effect of the Fe excess on the temperature dependence of the $B_{1g}$ mode linewidth in Se-substituted samples.}
\label{Fig4}
\end{figure}

We now discuss the effect of the excess iron on the vibrational properties for the Se-rich compounds. As seen directly by comparing the data in Figs.~\ref{Fig1}-b and -c, the low temperature behavior of the phonons is strongly affected by the Fe concentration.
In the case of Fe$_{0.95}$Te$_{0.56}$Se$_{0.44}$, as already discussed, regular hardening and narrowing of the phonon are observed as temperature is lowered. This is true between room temperature and 60 K for the Fe$_{1.05}$Te$_{0.58}$Se$_{0.42}$ sample, with a very close Se content to those of Fe$_{0.95}$Te$_{0.56}$Se$_{0.44}$ but where $\sim$ 5\% of excess Fe has been measured. A sudden upturn in this behavior is observed for lower temperature, since at 5K the mode is softer and broader than at 60K.
This can be better seen on Figs~\ref{Fig4}-a and b, where we compare the temperature dependence of frequencies and linewidth of the $B_{1g}$ mode for these two samples. Clearly, as temperature decreases, the phonon hardens and narrows regularly, just as in absence of excess iron, but at a temperature of about 35 K the mode starts to broaden ($\sim$ 3 cm$^{-1}$) and softens ($\sim$ 5 cm$^{-1}$).
In samples with lower Se contents a similar, although weaker, softening of the phonon is observed in the same temperature range, but with no change in the linewidth. None of these effects were observed on the $A_{1g}$ phonon (not shown here).

\subsection{Summary of the experimental results}
To summarize our main experimental findings, we observe the following:
\begin{enumerate}[label=\roman{*})]
 \item{For null or low Se concentration, an unusual broadening of the $B_{1g}$ phonon FWHM as temperature is decreased. This progressively turns into a conventional anharmonic narrowing as Se is substituted to Te.}
\item{Clear softening and narrowing of this $B_{1g}$ phonon through the magnetic transition in the parent compound, suppressed in presence of Fe excess.}
\item{Strong softening and broadening of the $B_{1g}$ mode at low temperature in Se substituted samples.}
\end{enumerate}

%In conclusion, in both parent and Se-substituted samples, the $B_{1g}$ broadens at low temperature if the iron concentration is increased.
%The effect of excess iron on the phonon energy is however different in the two cases, as a weak hardening (softening) is observed in the parent (Se-substituted) compounds, respectively.

\section{LDA DFT Calculations}
\label{DFTcalc}

In order to gain further insights on the effects of magnetism and Fe concentration on the lattice dynamics in these systems, we have performed DFT calculations of the frequency of the $A_{1g}$ and $B_{1g}$ modes in both paramagnetic (non-spin-polarized, nsp hereafter) and double stripe ordered (spin-polarized, sp) phases.
We start with the results obtained for stoichiometric FeTe system using the frozen phonon approach, summarized in Table~\ref{tab:TableDFe$_{1.02}$Te}.~\cite{DFT:details} Consistently with previous calculations,~\cite{Subedi08, singh2} we find a value of the double-stripe magnetic moment at $y=0$ (m =$ 2.2 \mu_B$), which is close to the experimental one.

\begin{table}[h]
\caption{\label{tab:TableDFe$_{1.02}$Te} $A_{1g}$ and $B_{1g}$ frequencies of FeTe, from
 non-spin-polarized ($\omega_{nsp}$) and spin-polarized ($\omega_{sp}$)
DFT calculations.}
\begin{ruledtabular}
\begin{tabular}{cccc}
mode  & $\omega_{nsp}$ (cm$^{-1}$) & $\omega_{sp}$ (cm$^{-1}$) & exp. at 10 K (cm$^{-1}$) \\
  & & & Fe$_{1.02}$Te sample \\
\hline
$A_{1g}$ &    135 &        175 & 159.7 \\
$B_{1g}$      &    200.9 &     197.5 & 200.5\\
\end{tabular}
\end{ruledtabular}
\end{table}

We then consider the effect of Fe excess and deficiencies on the $B_{1g}$ mode. Non-stoichiometry was considered within the VCA, which amounts to
 replacing Fe with a ``virtual'' atom with charge $Z_{Fe}^{\pm y}$ in the self-consistent DFT calculations.
This means that the  average potential due to doping is treated self-consistently, but the effects of randomness are disregarded.
Since, within VCA, it is  impossible to model the isovalent Se/Te substitution, we do not address the issue of the dependence of the $A_{1g}$ mode on Se/Te concentration in this work.
A proper description would require large supercell calculations, beyond the scope of the present paper. We used the FeTe experimental lattice parameters and Te height for all Fe concentrations.~\cite{Bendele10}
For each value of the Fe excess $y$, the frequencies were calculated for nsp and sp configurations.~\cite{DFT:details} The frequencies we obtained are given in Table~\ref{tab:TableDFT}, together with the self-consistent value of the magnetic moment at equilibrium.
%
%# sample      nsp (cm-1)     sp (cm-1)  mom Fe at zero displacement (mu_B)
\begin{table}
\caption{\label{tab:TableDFT}$B_{1g}$ frequencies of Fe$_{1+y}$Te, from
 non-spin-polarized ($\omega_{nsp}$) and spin-polarized ($\omega_{sp}$)
DFT calculations. $m$ is the value of the self-consistent
double-stripe moment at zero displacement, in $\mu_B$.}
\begin{ruledtabular}
\begin{tabular}{p{2cm}ccc}
  & $\omega_{nsp}$ (cm$^{-1}$) & $\omega_{sp}$ (cm$^{-1}$)& $m$ ($\mu_B$) \\
\hline
Fe$_{0.98}$Te  &    216.76 &        207.36 &    2.34\\
Fe$_{0.99}$Te  &    207.21 &        199.92 &  2.28\\
FeTe      &    200.94 &        197.54 &    2.20\\
Fe$_{1.02}$Te  &    191.30 &        192.40 &    2.06\\
Fe$_{1.06}$Te  &    173.00 &        182.00 &    1.60\\
\end{tabular}
\end{ruledtabular}
\end{table}

The values of the calculated frequencies are extremely sensitive to the Fe excess content, which also affects strongly the magnetic moment.
In the VCA, the magnetic moments and frequencies decrease monotonically with increasing Fe content, up to the highest doping we calculated ($y = 0.06$).
We find an almost linear increase of the $B_{1g}$ frequency with doping, with a linear slope of $523$ and $292$ cm$^{-1}$/$y$ for nsp and sp calculations, respectively.
As discussed in the next section, the experimentally observed effects of Fe non stoichiometry are much weaker.

\section{Discussion}

\subsection{Phonon Renormalization through $T_N$: comparison with other families}
\label{throughTN}
In parent Fe$_{1.02}$Te, we have observed a clear softening and a narrowing of the $B_{1g}$ phonon through the magnetic ordering transition at $T_N$.
Such narrowing has also been reported in parent BaFe$_2$As$_2$~\cite{Chauviere09, Rahlenbeck09} and CaFe$_2$As$_2$~\cite{Choi10} through the spin-density wave (SDW) transition. In the latter case only, it is accompanied by a jump of the phonon frequency that can, at least partially, be explained by the sudden collapse of the unit cell along the $c$-axis through the transition (such collapse does not take place in BaFe$_2$As$_2$).

On the contrary, in Fe$_{1+y}$Te, the $c$-axis lattice parameter has been found to expand slightly through the coupled structural-magnetic transition.~\cite{Martinelli_PRB2010} This certainly favors the observed softening, although one would in this case expect an abrupt jump in the phonon frequency at $T_N$ (the structural transition is first-order) rather than the observed smooth softening between $T_N$ and 10 K.

Regarding the agreement with the calculated phonon frequencies, it has been shown for the 122 and 1111 systems that the experimental phonon frequencies are closer from those obtained with a magnetic calculation than from those where magnetism is not included, even in the paramagnetic state.~\cite{Boeri_PRB2010, Hanh_PRB2009, Fukuda_PRB2011, Reznik09}
This is also the case here as we get a better match with our experimental data using the sp calculation rather than the nsp one (see Sect.~\ref{DFTcalc}).

Finally, as the phonon linewidth is inversely proportional to the lifetime of the excitation, its renormalization through $T_N$ reflects the changes in the coupling of the phonons to some decay channels.
In the 122 arsenides, as the SDW gap opens,~\cite{Akrap_PRB2009, Boris_PRL2009} a significant reduction of the electronic density of states at the Fermi level occurs, leading to a decrease of electron-phonon coupling that reasonably accounts for the observed sharpening of the phonons.~\cite{Rahlenbeck09}
Such an SDW gap opening has not yet been reported in the 11 compounds,~\cite{Chen_PRB2009, Xia_PRL09} but has recently been observed in ARPES experiments~\cite{ARPES_gap} and is very likely responsible for the $B_{1g}$ mode narrowing in the Fe$_{1.02}$Te sample.

\subsection{Se-substituted systems}
\subsubsection{Absence of superconductivity induced anomalies}
As mentioned in the Introduction, the $B_{1g}$ Fe mode that shows here the most striking doping dependence is active in all the families of Fe-based superconductors.
The absence of renormalization of this phonon as well as of the $A_{1g}$ one through the superconducting transition in the doped compounds is consistent with experiments carried out on the doped 122 arsenide compounds~\cite{Rahlenbeck09} (with the notable exception being the electron-doped Pr$_{x}$Ca$_{1-x}$Fe$_2$As$_2$,~\cite{Litvinchuk_PRB2011} where a small hardening of the $B_{1g}$ phonon through $T_c$ has been observed) or on the 111 compound LiFeAs~\cite{Um_PRB2012}, where no signature of the superconducting transition is seen either in the frequency or in the linewidth of these two modes. This is due to the fact that the superconducting gap amplitude is smaller than the phonon frequencies, hence his opening let them unaffected. The situation is similar for the 11 compounds, as the phonon frequency is much larger than the energies reported for the superconducting gap(s) in various experiments [2$\Delta \sim$ 2~meV (17 cm$^{-1}$)~\cite{Noat_JPCS2010}, 3.4~meV (27.4 cm$^{-1}$)~\cite{Hanaguri_Science2010} or 4.6~meV (37 cm$^{-1}$)~\cite{Kato_PRB2009} from STM, 2$\Delta \sim$ 3~meV (24 cm$^{-1}$) from NMR,~\cite{Arcon_PRB2010} and 2$\Delta \sim$6~meV (48 cm$^{-1}$)from specific heat~\cite{Klein_PRB2010}].

\subsubsection{Se-substitution induced evolution of the $B_{1g}$ mode FWHM temperature dependence}
\label{disc}
We now turn to the influence of Se concentration on the behavior of the $B_{1g}$ phonon.
As it is substituted into the system, we observe a weak increase of the $B_{1g}$ phonon frequency at the lowest temperatures (see Figs.~\ref{Fig_B1g} b to e), that goes along with the reduction of the $c$-axis parameter reported from x-ray and neutron diffraction experiments.~\cite{Yeh_EPL2008, Li09, Martinelli_PRB2010}
The room temperature linewidth for this mode is only weakly dependent on the Se concentration, although the FWHM clearly tends to sharpen with increasing Se content. This fact is already surprising as one may have rather expected the Se-substitution induced disorder to favor a broadening of the phonon lineshape instead of a narrowing. The normal state temperature dependence of the mode FWHM for different doping levels is even more puzzling. As noted in Sec.~\ref{ResultsSe}, the situation appears conventional in the Fe$_{0.95}$Te$_{0.56}$Se$_{0.44}$ samples close to half doping, with a continuous narrowing of the phonon with decreasing temperature.
On the contrary, as Se contents are reduced the phonon broadens with decreasing temperature. This is particulary clear for the two Se-free samples (Fe$_{1.02}$Te and Fe$_{1.09}$Te), as seen in Fig.~6-d.

The temperature dependence of the three samples with the highest Se content have successfully been fitted using a conventional anharmonic decay model (Eq.~\ref{e2}). Interestingly, looking at the fitting parameters of Table~\ref{tab:TableFitB1g}, the residual - or temperature-independent - linewidth $\Gamma_0$ (3.4 cm$^{-1}$) in Fe$_{0.95}$Te$_{0.56}$Se$_{0.44}$ is comparable with the one of the temperature dependent parameter $\Gamma$ (2.1 cm$^{-1}$), whereas
in Fe$_{0.98}$Te$_{0.66}$Se$_{0.34}$  and Fe$_{0.99}$Te$_{0.69}$Se$_{0.31}$ we have $\Gamma_0 \sim$ 10 cm$^{-1} \gg \Gamma$.

The situation is thus much closer to the one reported for Pr$_{x}$Ca$_{1-x}$Fe$_2$As$_2$~\cite{Litvinchuk_PRB2011} than for LiFeAs, where the residual linewidth was always vanishingly small.~\cite{Um_PRB2012} This shows that in Fe$_{1+y}$Te$_{1-x}$Se$_x$ systems, the contribution of anharmonicity to the $B_{1g}$ phonon lifetime is not the dominant one.
An additional decay channel for this phonon must therefore take over the usual anharmonic effects, and its contribution increases strongly as Se content is decreased.
Such behavior can have at least two possible origins: electron-phonon coupling and spin-phonon coupling.
In the first scenario, the increasing relative weight %of the temperature independent linewidth
$\Gamma_0$ with decreasing Se content can be trivially related to the increase of the electronic density of states at the Fermi level $N(E_F)$~\cite{Subedi08} [as $\Gamma_0 \propto N(E_F)$)].
This may, however, not be sufficient to account for the reported effect: Electron-phonon coupling is, in principle, temperature independent and can, therefore, hardly explain the increasing linewidth of the $B_{1g}$ phonon with decreasing temperature in the systems with the lowest Se contents (Fe$_{1.02}$Te, Fe$_{1.00}$Te$_{0.78}$Se$_{0.22}$).
Having in mind the increasing weight of magnetic excitations as Se concentration decreases toward the parent compound,~\cite{Wen09} we are naturally led to suggest that spin-phonon coupling may be the additional decay channel for the phonons.
The effects of excess Fe discussed in the next section indeed confirm an interplay between the lattice and spin degrees of freedom in these systems.

\subsection{Influence of the iron concentration}

\subsubsection{Comparison of experimental data with LDA DFT calculation}
According to our calculation in Sec.~\ref{DFTcalc}, increasing the Fe concentration induces a softening of the $B_{1g}$ mode frequencies in both sp and nsp calculations. The softening rates are found to be $523$ and $292$ cm$^{-1}$/$y$ for nsp and sp, respectively.
Experimentally, at low temperature, a small hardening is instead observed in the parent compounds, when going from Fe$_{1.02}$Te to Fe$_{1.09}$Te.%, in the parent compound.%, while the rather calculation predicts a softening.
~In Se-substituted samples on the other hand, a $\sim$7 cm$^{-1}$ softening between Fe$_{0.95}$Te$_{0.56}$Se$_{0.44}$ and Fe$_{1.05}$Te$_{0.58}$Se$_{0.42}$ $B_{1g}$ phonon frequencies is indeed observed ($\sim$70 cm$^{-1}$/$y$).
An effect of comparable size is found when comparing the $B_{1g}$ peak frequencies of Fe$_{0.95}$Te$_{0.56}$Se$_{0.44}$ ($\omega_{B_{1g}} = 205.6 cm^{-1}$) and Fe$_{0.98}$Te$_{0.66}$Se$_{0.34}$ ($\omega_{B_{1g}} = 203.8 cm^{-1}$), with significantly different Fe deficiencies (even though the Se concentrations in this case slightly differ). This gives a softening rate of $\sim$ 66 cm$^{-1}$/$y$.

In any case, the experimentally observed effects of Fe nonstoichiometry are much weaker than those theoretically calculated, even considering the sp calculation where they are the smallest. One has to keep in mind that accurate comparison to the calculation is tough as the presence of excess ($y > 0$) Fe complicates the situation. In fact, in the VCA there is no qualitative difference between Fe excess and deficiency, as the excess (deficient) charge are both
located around the Fe site. Experimentally, however, it is known that the excess Fe ions are located in the Te planes, and
this has qualitatively different effects, the most important being that the effective Fe magnetic moment is enhanced and not reduced,
due to the formation of local moments on the excess Fe in the Te planes. This cannot be taken into account by the VCA approach of treating the doping in our LDA calculations. Furthermore, it is important to point out that mean-field LDA DFT calculations cannot reproduce the renormalization of the $B_{1g}$ frequency and linewidth observed at low temperature in the Fe-rich systems, which are discussed in the next paragraph.

\subsubsection{Excess Fe-induced magnetic fluctuation}
We have seen in Sec.~\ref{Exp_Fe_undoped} that in the Fe$_{1.09}$Te crystal, we are not able to observe clearly the effect of the magnetic transition on the phonons. A small softening has been seen (see Fig.~\ref{Fig3b}-d), but no narrowing of the line shape.
In the Se-substituted system, as shown in Fig.~\ref{Fig4}, excess iron, in addition to a decrease in $T_c$, induces large effects on both frequency and lineshape of the phonons. The strongest effect occurs close to 50\% Se content, with a large softening and broadening of the $B_{1g}$ phonon below $T\sim$35 K, well above $T_c$. To our knowledge, no phase transition has been reported in this temperature range for this doping level, but the occurrence of short-range magnetic fluctuations has been reported.~\cite{Khasanov09}
In the undoped case with low excess iron concentration, it has been shown that a low energy gap in the spin-wave excitation spectrum opens when entering the magnetic state.~\cite{Stock} Increasing the excess iron concentration, this gap is filled up with low-energy spin fluctuations.~\cite{Stock}

In both doped and undoped cases, one effect of excess iron is, therefore, to induce low energy magnetic fluctuations in a temperature range at which we also observe a relative broadening of the $B_{1g}$ phonon, \textit{i.~e.}, a decrease of its lifetime.
This reinforces the point we made at the end of Sec.~\ref{disc}, indicating that the additional damping for the $B_{1g}$ mode may actually originate from its coupling to magnetic excitations. %As Furthermore, in the undoped case these excitations persists above $T_N$ ~\cite{Stock}, and could therefore explain the anomalous broadening of the mode observed in the normal state of Fe$_{1.02}$Te and Fe$_{1.09}$Te samples.

%%%%%%%%%%%%%%%%%%%%%%%%%%%%%%%CONCLUSIONS %%%%%%%%%%%%%%%%%%%%%%%%%%%%%%%%%%%%%%%%%%%
%%%%%%%%%%%%%%%%%%%%%%%%%%%%%%%%%%%%%%%%%%%%%%%%%%%%%%%%%%%%%%%%%%%%%%%%%%%%%%%%%%%%%%
\section{Conclusions}
We have carried out a systematic study of the lattice dynamics in the Fe$_{1+y}$Te$_{1-x}$Se$_x$ system, focusing more particularly on the $c$-axis polarized Fe $B_{1g}$ mode. In parent compounds, unlike other systems such as BaFe$_2$As$_2$ or LiFeAs, a nonconventional broadening of this mode is observed as temperature decreases, and a clear signature of the SDW gap opening is observed.
As Se is substituted to Te, the temperature dependence of this modes smoothly evolves toward a more regular situation, with the $B_{1g}$ phonon showing conventional anharmonic decay. A good agreement between the observed phonon frequencies and a first-principles calculation including the effects of magnetic ordering is found.
The temperature dependence of the phonon linewidth, as well as the effects induced by the Fe nonstoichiometry in these compounds, revealed a peculiar coupling of this mode to magnetic fluctuations in the Fe$_{1+y}$Te$_{1-x}$Se$_x$ system and can, to date, not be satisfactorily reproduced within  state-of-the-art computational approaches.

%%%%%%%%%%%%%%%%%%%%%%%%%%%%%%%%%%%%%%%%%%%%%%%%%%%%%%%%%%%%%%%%%%%%%%%%%%%%%%%%%%%%
%%%%%%%%%%%%%%%%%%%%%%%%%%%%%%%%%%%%Acknowledgements%%%%%%%%%%%%%%%%%%%%%%%%%%%%%%%
\section{Acknowledgement}

We thank A. Schulz for technical support and A.C. Walters for useful suggestions. This work has been supported by the European project SOPRANO (Grant No. PITN-GA-2008-214040), by the French National Research Agency, (Grant No. ANR-09-Blanc-0211 SupraTetrafer), and by the UK Engineering and Physical Sciences Research Council (MJR, EP/C511794).
%%%%%%%%%%%%%%%%%%%%%%%%%%%%%%%%%%%%%%%%%%%


\begin{thebibliography} {99}
\bibitem{DFT:details} We used the LDA approximation for the exchange
  correlation functional, and the linearly-augmented plane wave (LAPW)
  basis set with local orbitals. The $\mathbf{k}$-point integration
  was performed on a 4 8 4 grid for the doubled unit cell. The
  experimental lattice parameters are $a$=3.826\AA and $c$=6.273\AA,
  with $z_{Te}$=0.28141.
\end{thebibliography}
\end{document}